\documentclass[12pt]{iopart}
\usepackage{url}
\usepackage{subfig}
\usepackage{caption}
\usepackage{float}
\usepackage{subfloat}
\usepackage{cite}
\usepackage{graphicx}
\usepackage{xcolor}
\usepackage{multirow}
\usepackage[normalem]{ulem}
\restylefloat{table}
\usepackage{array}
\expandafter\let\csname equation*\endcsname\relax
\newcolumntype{P}[1]{>{\centering\arraybackslash}p{#1}}

\expandafter\let\csname endequation*\endcsname\relax

\usepackage{amsmath}

\newcommand{\matrx}[1]{{\mathbf{\textsf{#1}}}}
\newcommand{\adj}{\matrx{A}}
\newcommand{\dist}{\matrx{W}}

\begin{document}

\title{
Effects of syndication network on specialisation and performance of venture capital firms} 

\author{Qing Yao$^{1,4}$, Shaodong Ma$^2$, Jing Liang$^2$, Kim Christensen$^{4}$, Wanru Jing$^3$, Ruiqi Li$^{2,*}$}

\address{$^1$ School of Systems Science, Beijing Normal University, Beijing 100875, China\\
$^2$ UrbanNet Lab, College of Information Science and Technology, Beijing University of Chemical Technology, Beijing 100029, China\\
$^3$ E-link Ventures, Shanghai 200050, China\\  
$^4$ Blackett Laboratory \& Centre for Complexity Science, Imperial College London, London SW7 2AZ, UK\\
$*$ Author to whom any correspondence should be addressed.}
\ead{lir@buct.edu.cn}

\vspace{10pt}

\begin{abstract}
The Chinese venture capital (VC) market is a young and rapidly expanding financial subsector. Gaining a deeper understanding of the investment behaviours of VC firms is crucial for the development of a more sustainable and healthier market and economy. Contrasting evidence supports that either specialisation or diversification helps to achieve a better investment performance. However, the impact of the syndication network is overlooked. Syndication network has a great influence on the propagation of information and trust. By exploiting an authoritative VC dataset of thirty-five-year investment information in China, we construct a joint-investment network of VC firms and analyse the effects of syndication and diversification on specialisation and investment performance. There is a clear correlation between the syndication network degree and specialisation level of VC firms, which implies that the well-connected VC firms are diversified. More connections generally bring about more information or other resources, and VC firms are more likely to enter a new stage or industry with some new co-investing VC firms when compared to a randomised null model.
Moreover, autocorrelation analysis of both specialisation and success rate on the syndication network indicates that clustering of similar VC firms is roughly limited to the secondary neighbourhood. When analysing local clustering patterns, we discover that, contrary to popular beliefs, there is no apparent successful club of investors. In contrast, investors with low success rates are more likely to cluster. Our discoveries enrich the understanding of VC investment behaviours and can assist policymakers in designing better strategies to promote the development of the VC industry. 
\end{abstract}

\section{Introduction}
Venture capital (VC) has been an essential source of financing for start-up companies deemed to have a high growth potential with innovative technologies or business models \cite{sun2019venture,rothwell2013patenting,kortum2001does}. In the past a few decades, we have witnessed an enormous increase in VC investment activities and start-ups supported by VC, including Tesla, Amazon, Apple, Facebook, Google, Alibaba, and Whole Foods, which have significantly impacted domestic and global economies. Gaining a deeper understanding of investment behaviours of VC firms are crucial for the more sustainable and healthier development of the industry and economy \cite{li2021assessing,gu2022unveiling}.

On the one hand, given the ultimately limited knowledge of industries or investment stages and resources of individual entities, being specialised is a natural choice to obtain a good financial performance for VC firms \cite{gompers2009specialization}. Previous evidence indicates that specialised VC firms tend to outperform generalised ones on the quality of capital allocation across industries, as specialists may be more responsive to signals of specific industries \cite{gompers2009specialization}. Specialists can diversify their portfolios to become generalist. However, generalised VC firms may encounter difficulties in redeploying capital into industries with better opportunities~\cite{rajan2000cost,scharfstein1998dark,scharfstein2000dark}. 


On the other hand, previous literature also suggests that the role of networking (which can lead to diversification) is crucial for the investment activities and performance of VC firms. 
Gompers \textit{et al.}~\cite{gompers2020venture} surveyed hundreds of VC firms on `How do venture capitalists make decisions?' and attributed the success more to the management team than to business-related characteristics (e.g., product or technology).  However, similar case studies are limited to a local view focusing on individual VC firms. It does not have a global perspective that appreciates the importance of interactions between entities, e.g., either between individuals in management teams in different companies or collaborations between VC firms \cite{2007whom,luo2016ERGM,bonaventura2020predicting}. 
Joint-investment relation between VC firms (i.e., investing in the same start-up company at the same time) form complex collective interaction patterns that influence the flow of information and trust \cite{burt2009holes,burt2022cooperation}.
Hochberg \textit{et al.}~\cite{2007whom} examined U.S.-based VC funds and found that better-networked VC firms experience significantly better fund performance, as indicated by a higher fraction of investments that are successfully exited through an initial public offering (IPO) or mergers and acquisitions (M\&A)~\cite{2007whom}, and concluded `whom you know matters'. 
A better networking with other VC firms can generally bring insightful information \cite{burt2009holes}, more contacts, more deal flows, and more resources \cite{bygrave1988structure,ahuja2000collaboration,katila2008swimming}. That means a better networked VC can benefit from mutual complementarity~\cite{uzzi1996sources} and diversity~\cite{podolny2001networks,manigart2002european}, which help the firm invest against uncertainties of market and policy, free ridings and opportunism behaviours~\cite{williamson1985,lerner1994syndication,2007whom,wilson1968theory,sorenson2001syndication,sah1986architecture,manigart2002european}. Additionally, a better networking will build a VC firm reputations~\cite{smith2010venture,hochberg2010networking,sathe2010ventureSecrects}.
This suggests that networking positively affects the investment performance, as VC firms can benefit from mutual knowledge and skill complementing from collaborators \cite{uzzi1996sources} and diversity of investments \cite{podolny2001networks,manigart2002european}.

There seems to be a fundamental contradiction between specialisation and generalisation in previous literature, thus, a comprehensive understanding of the relation between network effect on specialisation/generalisation (diversification) and investment performance needs to be developed. 
In this work, we exploit an authoritative dataset that records over $160,000$ detailed investment activities involving more than $31,000$ VC firms and $50,000$ start-up companies in the Chinese VC market since 1985. We first construct the syndication network of VC firms, where two VC firms are connected if they have made joint-investment in the same start-up company in one deal, i.e., at the same time. We reveal that better networked VC firms tend to have a more diversified spectrum of investments in terms of both investment stages and industries. In this project, we treat the co-investing VC firms as `friends' of each other, because, they have common interests and are connected through the same start-up companies. In addition, when entering a new industry, VC firms are more likely to collaborate with new friends, which might indicate that new friends bring about new knowledge, information, or other resources. 
When we analyse the correlation patterns of VC firms in the syndication network, we find a strong clustering effect that VC firms with similar specialisation levels tend to neighbour each other. Such clustering pattern is generally limited to the second order of neighbourhood. 
In addition, generalists are typically of a stronger local clustering, and the local clustering effect is weakening along with the increase of specialisation level. While for the investment performance, there is no obvious successful club where successful VC firms cluster together, but poor performance VC firms are surrounded by similar entities, that is, VC firms which perform poorly cluster together.






\section{Methods}

\subsection{Herfindahl Index for evaluating the degree of specialisation}
The Herfindahl Index, also known as Herfindahl--Hirschman Index (HHI), was first proposed to measure the competition between companies based on the market share of each firm~\cite{rhoades1993herfindahl}. 
With a similar formalisation, HHI later got adopted to evaluate the degree of specialisation of VC firms, which is the sum of the squares of the percentage of all previous investments of a VC firm in each industry \cite{gompers2009specialization} or in each investment stage. 
For example, if a VC firm $i$ (VC$_{i}$) invests in $\alpha$ industry with a fraction $p^{\alpha}_i = n_i^{\alpha}  / \sum_{\alpha=1}^{m_i} n_i^{\alpha}, \alpha = 1, 2, \dots, m_i $, where $n_{i}^{\alpha}$ is the number of investments the VC$_i$ makes in $\alpha$ industry, $m_i$ is the total number of industries VC$_i$ entered (which can be different across VC firms), and $\sum_{\alpha} n_i^{\alpha}$, is the total number of investments the VC$_i$ has made. The counting of investment numbers is illustrated in Fig~\ref{fig:fig1}(a). Then its HHI of industry, $H_i^\textrm{ind}$,  is 
\begin{equation} \label{eq.HHI}
    H_i^\textrm{ind} = \sum_{\alpha=1}^{m_i} \left(p_i^\alpha\right)^2.
\end{equation}
The most specialised VC firms ($m_{i} = 1$) have $H_i^\textrm{ind}=1$, and the most diversified or generalized ones have $H_i^\textrm{ind}=1/m_{i}$, which corresponds to the situation with equal number of investments $n^{\alpha}_{i}$ for each industry $\alpha = 1, 2, \dots, m_{i}$. The formalisation for the HHI of investment stage $H_i^\textrm{stage}$ is the same as Eq.(\ref{eq.HHI}) with $p^{\alpha}_{i}$ the fraction of investments in a certain stage $\alpha$.

\subsection{Moran's Index in networks}
The Moran's Index (Moran's I) is widely used to measure the spatial autocorrelation of concerned variables~\cite{cli1973spatial,cliff1981spatial,anselin1988spatial}, and here we adopt it to measure autocorrelation between the variables of VC firms in the syndication network. The Moran's I for a  variable with $N$ observations $x_{i}$ is defined as:
\begin{equation} \label{eq.GlobalMoran}
    I = \frac{\sum\limits_{i=1}^{N}\sum\limits_{j=1}^{N}w_{ij}(x_{i}-\bar{x})(x_{j}-\bar{x})}{|\dist|\sigma_{x}^{2} },
\end{equation}
where $N$ is the number of VC firms indexed by $i$; 
$x_i$ is the variable of interest of node $i$ (e.g., it can be HHI of stage or industry, or success rate of a VC firm); $\bar {x} = \frac{1}{N}\sum_{i=1}^{N}x_{i}$ is the mean of $x$; $\sigma_{x}^{2} = \sum\limits_{i=1}^{n}(x-x_{i})^2 / N $ is the variance of that variable. $w_{ij}$ is the spatial weight between nodes ${i}$ and ${j}$ and $w_{ii} = 0$. The sum is $|\dist| = \sum_{i=1}^N\sum_{j=1}^N w_{ij}$. We can define a spatial weight matrix $\dist$ with entries $w_{ij}$. Literature, especially in geosciences, use this measurement to identify the geographic difference in or characterise the regional variations of the variables of interest. It is a measurement that can combine relational information (i.e., how close the two entities are) and non-relational information (i.e., any individual performance of an entity/VC firm). 

A network's topological measurements are similar to the geometrical measurements, i.e., the shortest path length between two nodes characterises how `close' they are. Therefore, we can define Moran's I using network measurements to evaluate the topological position's impact on the non-network variables of interest, taking the performance of VC firms, for example. 

The local Moran's I of a node $i$ on networks in this paper is defined as:
\begin{subequations} \label{eq.LocalMoran}
\begin{align}
I_{i}^{local} &=\frac{N}{|\dist|}\frac {x_{i}-{\bar {x}}}{\sigma_{x}^{2}} \sum _{j=1}^{N}w_{ij}(x_{j}-{\bar {x}}),\\
I &= \frac{1}{N}\sum_{i=1}^N I_i^{local}.
\end{align}
\end{subequations}


To make further investigation on the effects of topological distance on the autocorrelation of variables, we define four different weight matrices $\dist$ based on different levels of neighbourhood: 
considering the nearest neighbours (i.e., first-order neighbourhood), further integrate the second-order, third-order, and all reachable neighbours, respectively. 
The detailed definitions are:
\begin{subequations}\label{eq:weight}
\begin{align}
\dist_{ij}^{(1)} &= \adj_{ij}^{(1)},\label{eq.adj1}\\ 
\dist_{ij}^{(2)} &= \adj_{ij}^{(1)} + \frac{1}{2}\adj_{ij}^{(2)},\label{eq.adj2}\\ 
\dist_{ij}^{(3)} &= \adj_{ij}^{(1)} + \frac{1}{2}\adj_{ij}^{(2)} + \frac{1}{3}\adj_{ij}^{(3)},\label{eq.adj3}\\
\dist_{ij}^{(L)} &= \adj_{ij}^{(1)} + \frac{1}{2}\adj_{ij}^{(2)} + \frac{1}{3}\adj_{ij}^{(3)} + \frac{1}{4}\adj_{ij}^{(4)} + \cdots+ \frac{1}{L}\adj_{ij}^{(L)},\label{eq.adjL}
\end{align}
\end{subequations}
where $L$ is the diameter of the network and 
\begin{equation} \label{eq:adj}
    \adj_{ij}^{(n)} = \left\{\begin{matrix}
1 & i \neq j \text{ and with a shortest path $l_{ij}=n$}\\ 
 0 & \text{otherwise}
\end{matrix}\right.
\end{equation}
represents the non-zero entry of $n$-th order of adjacency, for example, $\adj_{ij}^{(1)}$ is the ordinary adjacency matrix of the network, and $\adj_{ij}^{(2)}=1$ indicates there is at least one shortest path with a path length $l_{ij}=2$ between node $i$ and $j$. 


\section{Results}
\subsection{Fast growing Chinese VC market}

\begin{figure}    \centering
\includegraphics[scale=1]{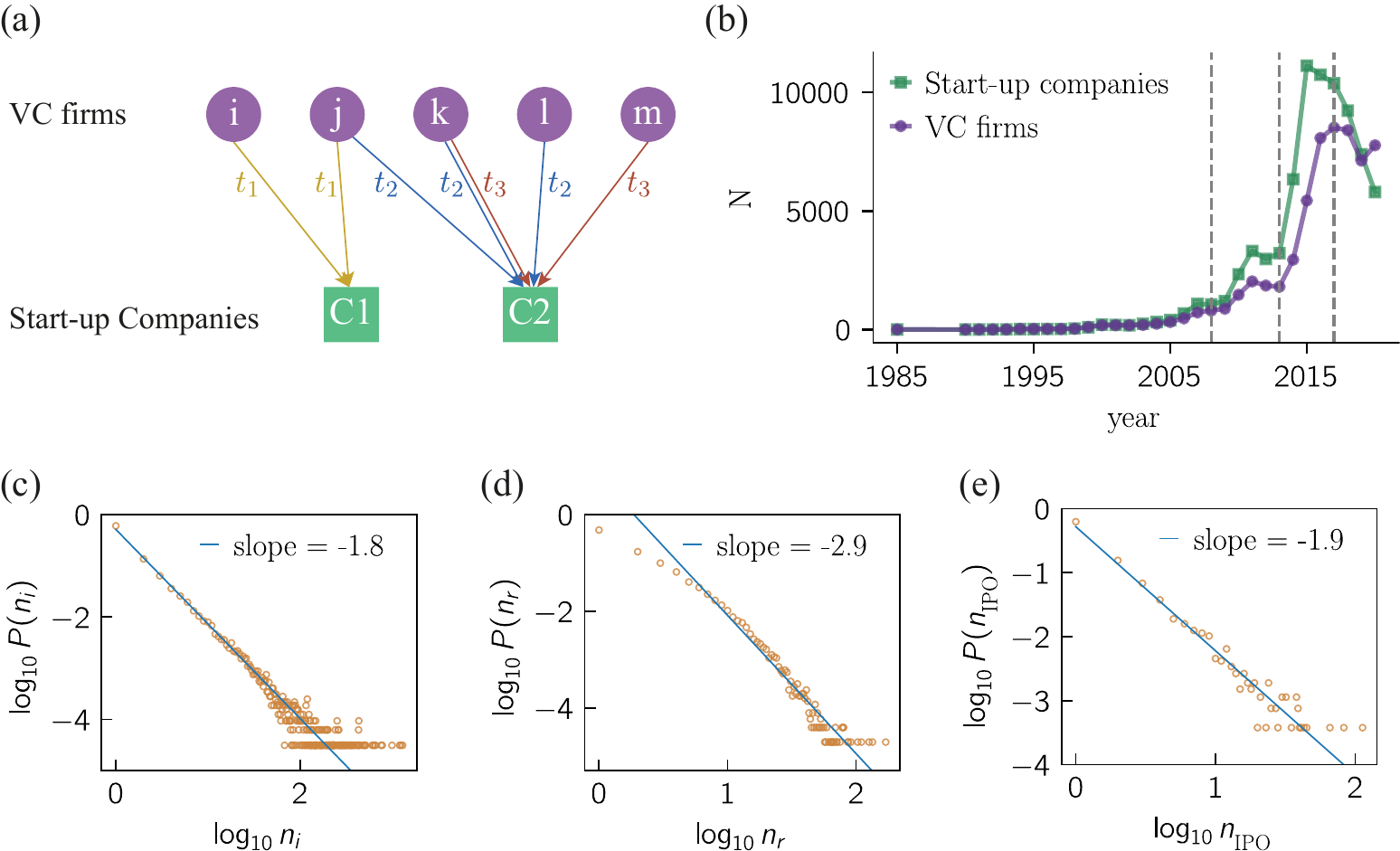}
\caption{(a) The illustration of investment relationships between VC firms and start-up companies. The purple circles represent VC firms while the green squares are the start-up companies. Every arrow associated with time represents  one investment, it can happen in the same or different deals (times). The number of investment is the number of arrows. For example, $n_{i}^{\texttt{C}_{1}} = 1$ and $n_{k}^{\texttt{C}_{2}} = 2$. The quantity includes different industries or stages. (b) The number of VC firms (purple circles) and the number of start-up companies (green squares) during the period 1985 - 2020. A sharp increase happened around 2013-2014. (c) The distribution of the number of investments made by each VC firm, i.e., the number of outgoing arrow $n_{i} = \sum_{\alpha = 1}^{m_{i}}n_{i}^{\alpha}$ of each purple circle. (d) The distribution of the number of investments received by each start-up company, i.e., the number of incoming arrow on each green square. (e) The distribution of the number of IPOs of each VC firm. The three distributions are all approximated by a power-law like distribution, indicating the heterogeneous characteristics of both the investments and the performance of the VC firms.}
\label{fig:fig1}
\end{figure}

The VC industry originated after World War II and matured in the late 1970s~\cite{2007whom}. There have been several boom-and-bust cycles in the VC industry decades ago in the western world~\cite{gompers2004venture}. However, VC is still a relatively newly emerging industry in China, where government policies keep changing, the governance structure is immature, and information asymmetry always bothers investors~\cite{luo2015,website:BankVC}. Nevertheless, the Chinese VC industry has undergone fast growth in the recent decades, especially after 2013 (see Fig.~\ref{fig:fig1}(b)). Although a drastic decrease in active VC firms happened after 2017, the number of start-up companies that receive VC investments remains high (see Fig.~\ref{fig:fig1}(b)). 

We split our dataset into two parts: 1985-2013 and 2014-2020. We use investment data before 2014 to construct syndication network, and use exit events data from 2014 to 2020 to make better evaluations on the outcome of previous investments. The duration of the second part is seven years and is long enough to observe the results of investments made before 2014. As a start-up company needs time to grow or to mature, VC investments generally requires several years to exit \cite{bonaventura2020predicting}. 


The number of investments of individual VC firms, $\sum_{\alpha=1}^{m_{i}}n_{i}^{\alpha}$ exhibits a power law (see Fig.~\ref{fig:fig1}(c)), which indicates that most VC firms only make a few investments, and only a notable small fraction of VC firms making many investments. However, since the slope $-1.8$ is larger than $-2$, the investments made by this small fraction of VC firms might take up the majority of VC investments in the whole market, because the average number of investments is determined by the tail of the distribution.
The number of VC investments received by each start-up company also exhibits a power-law tail, implying some start-up companies receive a much larger number of investments from VC firms (see Fig.~\ref{fig:fig1}(d)). 
However, the distribution of investments made by VC firms are more heterogeneous as indicated by a flatter exponent compared with Fig.~\ref{fig:fig1}(d), reflecting that the probability of a VC firm has a large number of investment activities is higher than the probability of a start-up company receiving lots of investments. 
Evaluating and predicting the financial investment performance is at the core of VC studies~\cite{2007whom,gompers2009specialization} but challenging. 
The best way to evaluate the financial performance of a VC firm is using the internal rate of return or return of investment. Unfortunately, such data is difficult to obtain due to business concerns. When investments exit through IPOs, M\&As or receiving the next round investment, VC firms make a great profit or have higher evaluation of the invested start-ups. Other types of exits are usually not successful~\cite{2007whom}. Therefore, we choose the number of IPOs, M\&As and receiving next round investment as the proxy for investment performance, following the works of~Refs.~\cite{2007whom,bonaventura2020predicting}.  We find that the number of IPOs of VC firms exhibits a power law (see Fig. \ref{fig:fig1}(e)). Similar behaviour is observed for the number of M\&As,  the number of investments that are not closed yet but receive a next round of investment, and unsuccessful exits (see Appendix Fig.\ref{fig:figA1}(a)-(c)). Again, these power-law like investment distributions indicate the heterogeneous nature of the VC market.

\subsection{Being specialised and in the joint-investment network}

Making investment decisions requires a versatile knowledge and resources depending on different industries and stages. Thus, being specialised is typically a sensible choice for VC firms to guarantee long-lasting and reproducible successful investments~\cite{rajan2000cost,gompers2009specialization}. 

According to the categories given by the Industrial Classification of National Economic Activities (ICNEA), there are seventeen industries in the first level classification, therefore, the range of the HHI of industry ($H^{ind}$) is within the interval $\left[1/17, 1 \right]$ 
(see the distributions of HHI of VC firms in Fig. \ref{fig.A:HHIdistribution} in Appendix). 

Except for the well-understood industry entry barriers~\cite{demsetz1982barriers}, we would like to further explain the four different investment stages of VC investments. In each stage, apart from providing capital support, VC firms need to assist the start-up on various aspects with different focuses. During the \textit{seed} stage, the VC investors focus on helping the start-up polish its products and build the team. In the \textit{initial} stage, dissecting market opportunities and finding the target markets would generally be the focus. In the \textit{expansion} stage, VC firms needs to help expand the market or diversify and differentiate product lines. 
And at last, in the \textit{mature} stage, VC firms prioritise how to go public. The four stages gives the range of HHI of stage ($H^{stage}$) $\left[1/4, 1 \right]$ 
(see the distributions of HHI of VC firms in Fig. \ref{fig.A:HHIdistribution} in Appendix).

\begin{figure}[!h]    
\centering
\includegraphics[scale=1]{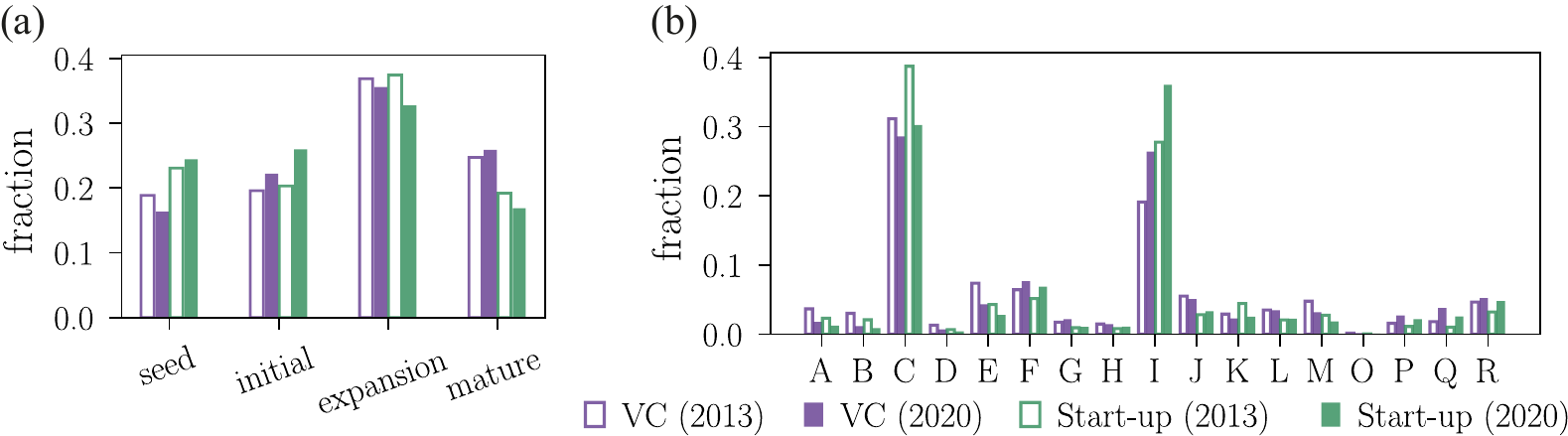}
\caption{The fraction of VC investments in each of the (a) $4$ investment stages and each of the (b) $17$ industries in 2013 (open) and 2020 (filled), where letters indicate the first level classification in the Industrial Classification of National Economic Activities (ICNEA). Even with the sharp increase in number of VC firms and start-up companies in 2013, the investment preferences remain very similarly: the investments in expansion stage is popular; two industries including `manufacturing' (C) and `information transmission, software and information technology service' (I) receive much more investments than any other industry.} 
\label{fig:fig2}
\end{figure}

Even though the Chinese VC market undergoes viral growth after 2013, the fraction of VC investments in 2020 in different investment stages (see Fig.~\ref{fig:fig2}(a)) and industries (see Fig.~\ref{fig:fig2}(b)) remains quite close to the 2013 snapshot. Most VC investments fall into the expansion stage and into `manufacturing' (code C of ICNEA) and `information transmission, software and information technology service' (code I of ICNEA), with a slight decrease in manufacturing and an increase in the information industry.



As mentioned in the Introduction, evidence shows that both specialisation and generalisation can be beneficial to investment performance~\cite{gompers2009specialization,2007whom,manigart2002european}. Why do we have this seemly contradictory phenomenon? We need new perspectives to comprehensively quantify the impacts of networking on specialisation and investment performance.  

\begin{figure}[!h]
\centering
\includegraphics[width=\textwidth]{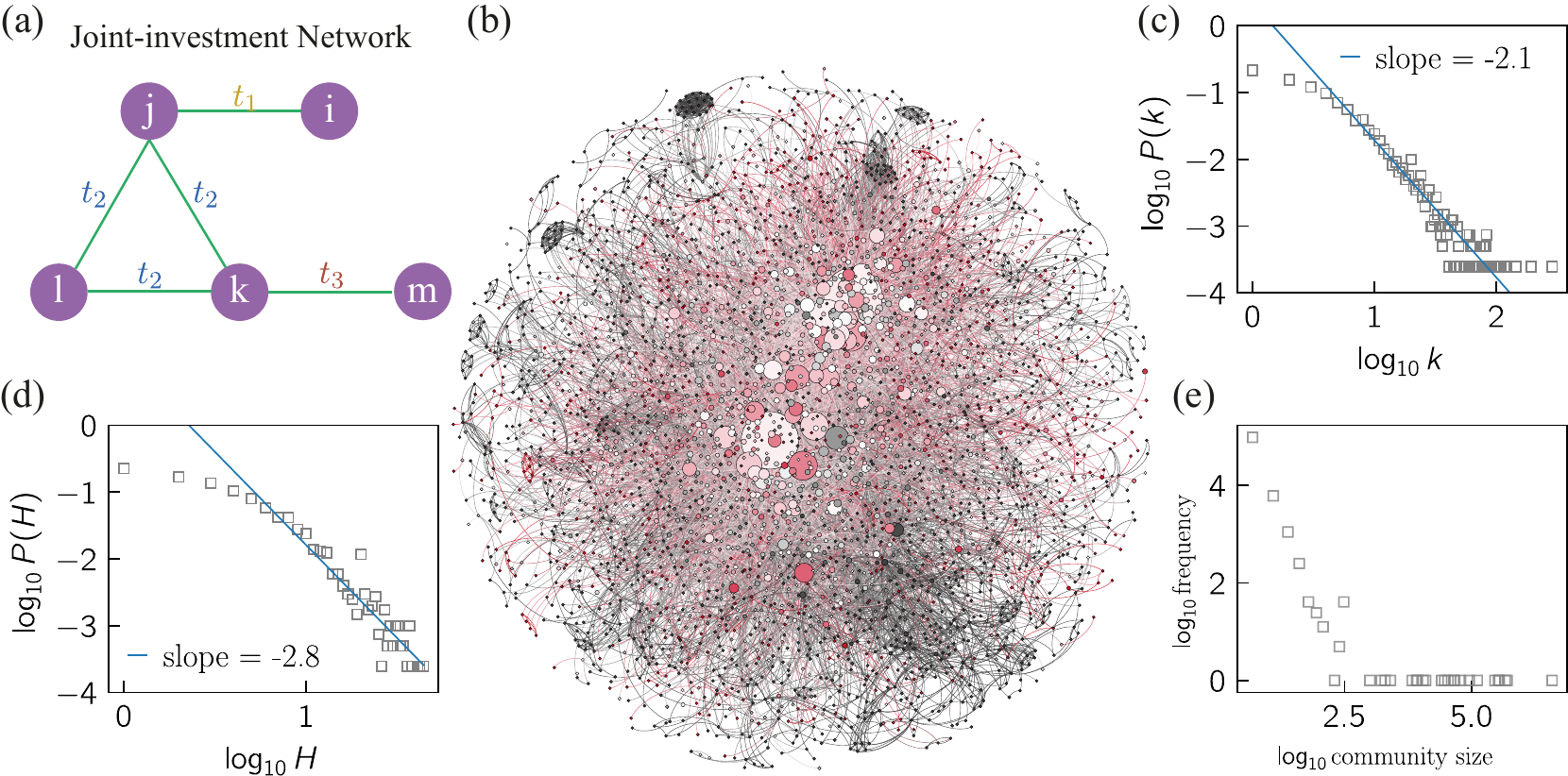}
\caption{(a) The illustration for the joint-investment network constructed from the history of investments in Fig~\ref{fig:fig1}(a). The purple circles represent VC firms and the VC firms are connected if they invest in the same start-up company at the same time. For example, node $j$, $k$ and $l$ are connected, because they all invest in start-up company C2 at $t_{2}$; whereas $m$ merely connects to $k$, as $k$ and $m$ co-invested in start-up company C2 at a later time $t_{3}$. (b) The joint-investment network of VC firms in 2013. The size of nodes indicates the number of investments, and the colour indicates the success rate, which is between zero (red) and one (black). The large node with many degrees are not with high success rate while some clusters not at the core are with high success rates. The frequency distribution of (c) degree $k$, and (d) H-index of VC firms. These power-law like distributions of network measurements emerge from the interactions among VC firms, implying not a random but interdependent mechanism of the joint-investment.  (e) The frequency distribution of community sizes in the log-log scale. The communities are detected by the Louvain method~\cite{blondel2008fast}. This plot indicates there exists a few large-sized and many small-sized communities.} 
\label{fig:nets}
\end{figure}

To study the network effects, we build the syndication network of VC firms based on their investment activities. In the syndication network, nodes are VC firms, and edges signify joint-investments in the same start-up company at the same time, see Fig~\ref{fig:fig1}(a) and Fig~\ref{fig:nets}(a).
The joint-investment network of VC firms in the Chinese market has grown quite dense with the emergence of many hub nodes that maybe of great importance (see Fig. \ref{fig:nets}(b)). Recent advance indicates that so called H-index can reveal the importance of a node in the network~\cite{lu2016h}. The node centrality H-index measure is defined to be the maximum value $H$ such that there exists at least $H$ (nearest) neighbours of degree larger than or equal to $H$. If a node has H-index $5$, it has $5$ neighbours with degree larger than or equal to $5$ but does not have $6$ neighbours with a degree larger than or equal to $6$~\cite{lu2016h}. We find that the distribution of the H-index also exhibits a power-law tail (see Fig.~\ref{fig:nets}(d)). There is also a profound community structure in the Chinese VC market (see Fig. \ref{fig:nets}(e)), where VC firms within the same community would have denser connections with one another than with others outside that community~\cite{blondel2008fast,liu2022revealing,shang2022local}. 



\subsection{Impacts of networking on specialisation}

\begin{figure}
\centering
\includegraphics[scale=1]{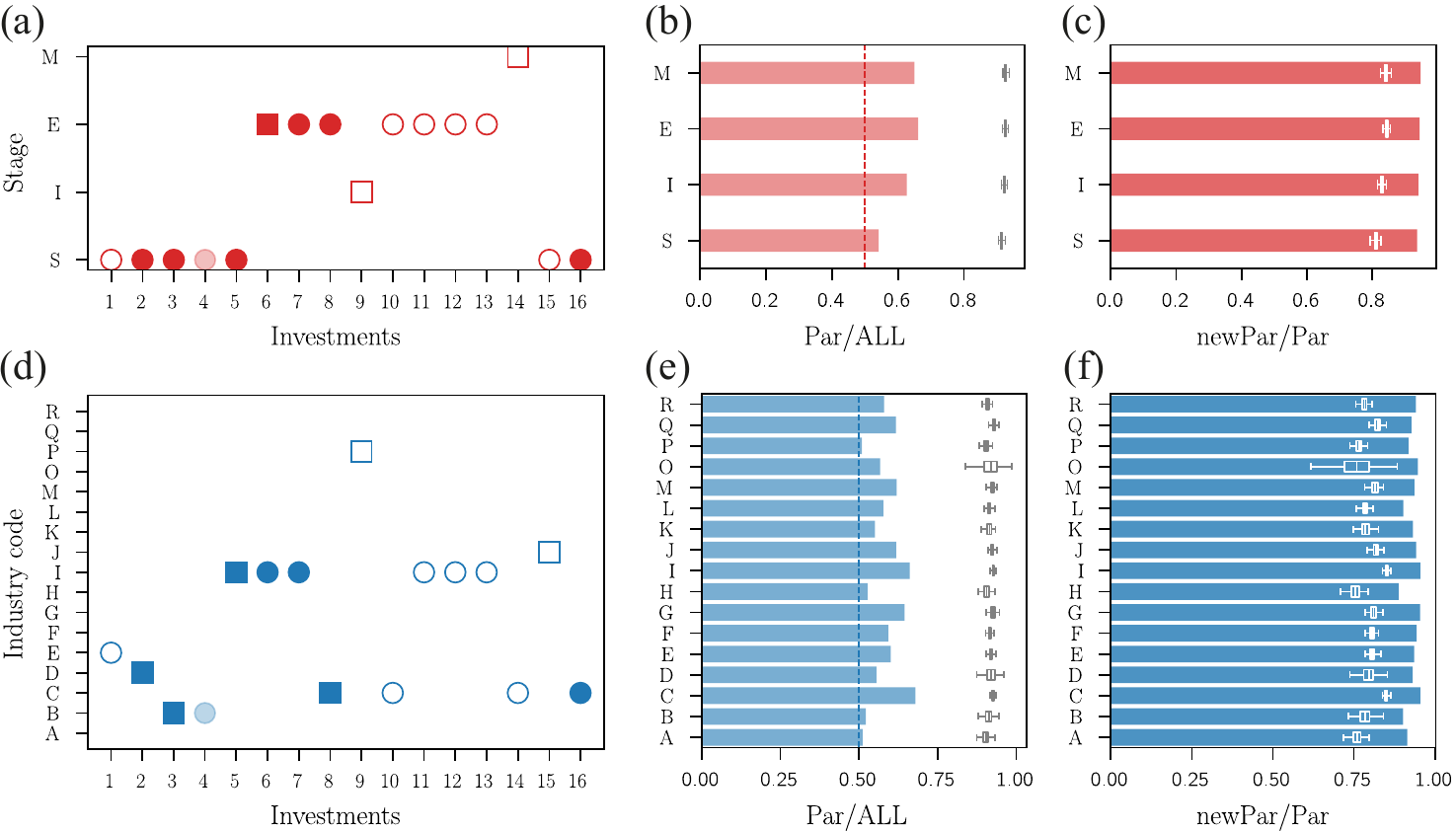}
\caption{The investment sequence illustration of an anonymous VC firm is categorised by (a) stage (red) and (d) industry (blue) of the start-up companies it invested in. That VC makes sixteen investments involving four investment stages and seven industries and all its investments are ordered by time. 
The investments made alone or joint-investment with others are denoted by open symbols and filled symbols, respectively. Additionally, the first investment entering either a new stage (see (a)) or a new industry (see (d)) is denoted by squares, while non-fresh trials are denoted by circles (a VC usually makes its first investment in the stage and industry which it is most familiar with, therefore, we do not treat it as a new stage/industry). When a VC firm makes joint-investments with old friends and no new friends, the filled marker is denoted with a lighter colour, which only happens once in all of its sixteen investments in the example shown in (a) and (d). 
The fraction of joint investments when a VC firm enters (b) a new stage or (e) a new industry. The fraction of making joint investments with a new friend (i.e., VC firms with no previous collaborations) when a VC enters (c) a new stage or (f) a new industry. 
Null models in (b-c) and (e-f) are all generated two hundred times.} 
\label{fig:HHIvsDegree}
\end{figure}

As suggested by previous literature \cite{2007whom,manigart2002european,uzzi1996sources}, for a VC firm, collaborating with others generally brings more information and knowledge, which is needed when entering an unfamiliar industry or stage. However, there is no quantitative analysis on the relation between networking (i.e., making joint-investments with others) and specialisation and, eventually, the performance of the VC firm. Here, we scrutinize the investment sequence of each VC firm to see if there is a higher probability for VC firms to enter a new stage or industry with others, especially with new friends who might have new knowledge, skills, or other resources. 

In Fig.~\ref{fig:HHIvsDegree}, we take an anonymous VC firm as an illustration. It enters three new stages (indicated by squares), one of which is joint-investment with others (indicated by filled symbols). 
We calculate the fraction of joint-investment out of all first-investment when entering a new industry for all VC firms, and find such a fraction is generally higher than 50\% (see Fig.~\ref{fig:HHIvsDegree}(b)).  
The same procedure is applied to analysis of investments when entering a new industry (see Fig.~\ref{fig:HHIvsDegree}(d)), and very similar results are obtained (see Fig. \ref{fig:HHIvsDegree}(f)). 
This suggests that when a VC firm enters either a new stage or industry, it would have a high chance of making collaboration. This phenomenon most likely occurs because VC firms employ collaboration to complement knowledge and skills against uncertainty. However, given the frequency and the size of collaboration, the fraction of joint investment is lower than the randomized null model, where we randomly assign each of their joint-investment partners to an investment event (see Table~\ref{tab:A.randomization} in Appendix for the randomisation process). 

Furthermore, for all joint-investments entering a new stage or industry, we analyse the fraction of joint-investments with new friends, i.e., the ones that have no collaboration before. The fractions are quite high, around 90\% for both entering a new stage and industry (see Fig. \ref{fig:HHIvsDegree}(c) and Fig. \ref{fig:HHIvsDegree}(f)), and it is also much higher than the random configurations in the null model. This indicates that collaborations with new friends generally happen with a denser concentration and suggests that there are certain mechanics other than randomness driving joint-investments with new friends when entering unfamiliar situations. One of the possible reason is that new friends might be crucial for bringing in new information and resources, another explanation is that `collaboration' with new friends is not intended, yet just a consequence of entering new stage or industry where the VC firm has few old friends. 

\begin{figure}
    \centering
    \includegraphics[scale=1]{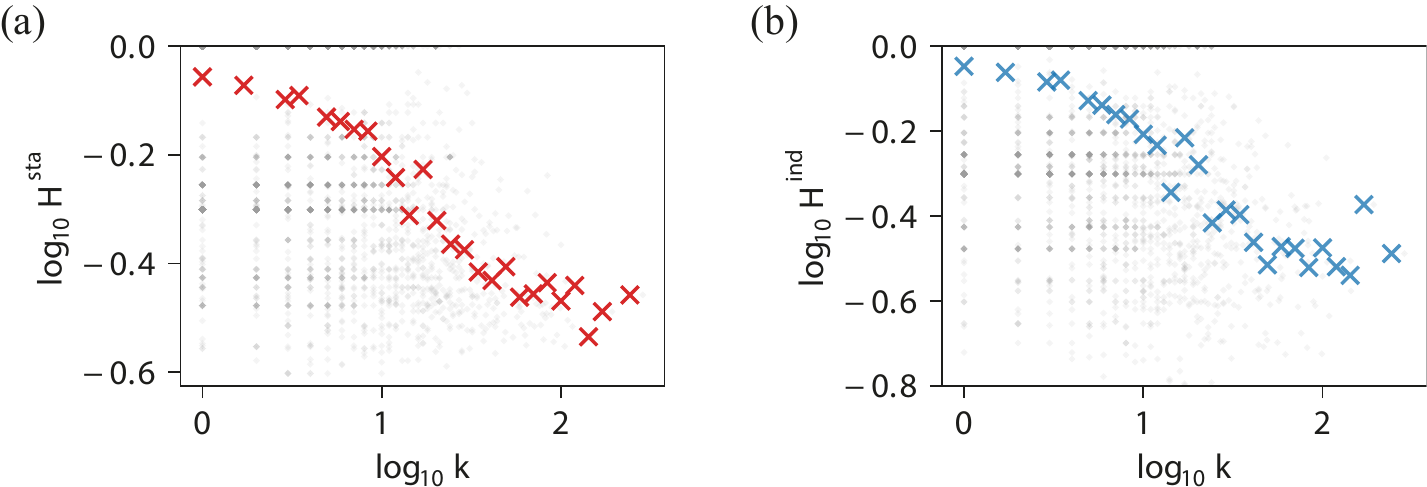}
    \caption{The relation between degree and specialization on (a) investment stages and (b) industry. For (a) and (b), we analyse VC firms with no less than five investments to guarantee a better statistical analysis as it is more convincing to evaluate the HHI of four stages based on at least five times investment activities.}
    \label{fig:Hvsk}
\end{figure}

By analysing the relationships between degree of the joint investment networks and the HHIs, we discover that when a VC firm has more friends (i.e., a larger degree in the syndication network), their level of specialisation on both investment stages (see Fig.~\ref{fig:Hvsk}(a)) and industries (see Fig.~\ref{fig:Hvsk}(b)) decrease (i.e., they are becoming more like a generalist), and the trends of deceasing become stronger when the degree is roughly beyond eight for both cases, which might be also fundamentally related to branching processes in networks \cite{evans2022linking}.

\subsection{Autocorrelation analysis on the joint-investment network}
Given the heterogeneous nature the joint-investment syndication network, to further quantify the relation between networking and specialisation and investment performance, 
we adapt Moran's I in geoscience (see Sec.2 Methods) to study correlation between network measurements and the non-network properties of VC firms. 
In spatial analysis, the level of adjacency (i.e., the weight matrix $\dist$ in Eq.(\ref{eq:weight})) between entities can be defined based on different Euclidean distance threshold \cite{cliff1981spatial} or other criteria \cite{wang2021revealing}; on networks, we can specify the different level of adjacency and 
we define four types of weight matrices. 
The first one is the ordinary adjacency matrix ($\dist_{ij}^{(1)}=\adj_{ij}$), i.e., if two VC firms have joint-investment, then $w_{ij}=1$, otherwise, $w_{ij} = 0$ (see Eq.(\ref{eq.adj1})). As information and trust can spread on networks, we further integrate the second-order neighbours ($\dist_{ij}^{(2)}$), but with a smaller weight for them: $w_{ij}=1/2$ if there exists a shortest path between $i$ and $j$ with $l_{ij}=2$ (see Eq.(\ref{eq.adj2})). We further accumulates third-order neighbours ($\dist_{ij}^{(3)}$), and all reachable neighbours ($\dist_{ij}^{(L)}$, where $L$ denotes the diameter of the network) in the same way, see Eq.(\ref{eq.adj3}) and Eq.(\ref{eq.adjL}), respectively.

\begin{table}[!htbp] \centering
\begin{tabular}{|P{5em}|P{5em}|P{5em}|P{5em}|}
\hline

& \textbf{$I^\textrm{stage}$} & \textbf{$I^\textrm{ind}$} & \textbf{$I^\textrm{success}$}  \\
\hline
 \boldmath $\dist_{ij}^{(1)}$        & $0.558$                & $0.557$                    & $0.286$                       \\
\hline
\boldmath $\dist_{ij}^{(2)}$        & $0.354$                 & $0.328$                    & $0.080$                        \\
\hline
\boldmath $\dist_{ij}^{(3)} $        & $0.077$                 & $0.074$                    & $0.021$                       \\
\hline
\boldmath $\dist_{ij}^{(L)}$   & $0.011$                 & $0.011$                    & $0.004$                       \\
\hline
\end{tabular} 
\caption{The global Moran's I at different level of adjacency. The measurements include HHI of stage, HHI of industry and success rate. For instance, $I^\textrm{stage}$ is calculated by substituting $H^\textrm{stage}$ as $x$ in Eq.(\ref{eq.GlobalMoran}). We permutate the values of the three measurements for $200$ times and calculate the $p$-value of the observed Moran's I. The $p$-values of all the measurements are $<0.001$, indicating the significance of the autocorrelation of the measurements of nodes and their neighbours. The decreasing trend of Moran's I as a function of distance is visualised in Figure~\ref{fig:morani_distance}.}
\label{tab:moran_wij}
\end{table}

\begin{figure}[!htbp]     
\centering
    \includegraphics[scale=1]{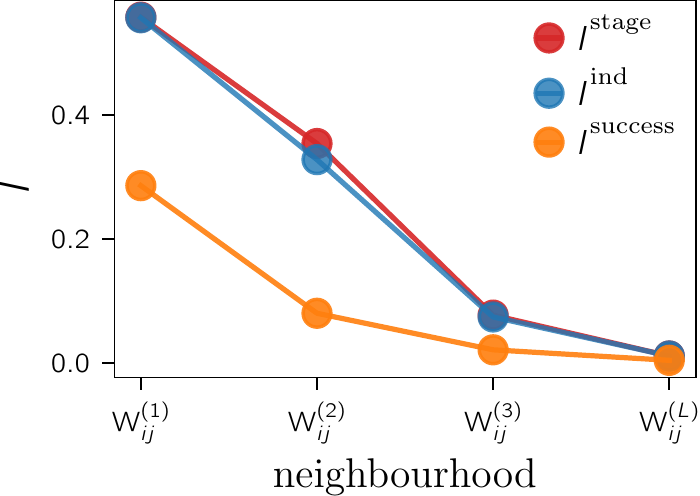}
    \caption{The relationship between the level of adjacency and the global Moran's I. For all three variables, the global Moran's I decreases with the level of adjacency. The Moran's I of success rate is always lower than both the stage and industry specialisation levels, which exhibit similar behaviours. $L$ refers to the diameter of the network, which corresponds to all nodes reachable in the network.}
    \label{fig:morani_distance}
\end{figure}

The global Moran's I for different weight matrices $\dist$ are summarised in Tab.~\ref{tab:moran_wij}. We find that for all measurements $p$-value $<0.001$, which demonstrate the significance of the local clustering of the concerned measurement (e.g., specialisation or performance). It is also worth noting that the global Moran's I for HHI of stage and industry are all high. 
Such clustering phenomenon suggests the trend that VC firms tend to make joint-investment with others alike. Moreover, there is a clear pattern of negative correlation between the global Moran's I and different level of adjacency (see Fig.~\ref{fig:morani_distance}). When considering the weighted adjacency with three-order neighbours, the global Moran's I becomes much smaller for all three measurements ($H^{ind}$, $H^{stage}$, and success rate), which suggest that at a larger range, the autocorrelation is much weaker, and the clustering effect is only obvious within second-order (e.g., for $H^{ind}$ and $H^{stage}$) or even just first-order (e.g., for success rate) neighborhood (see Fig.~\ref{fig:morani_distance}). Such a discovery is also consistent with previous findings on VC firm syndication patterns that the syndication probability drops quite fast when the distance between two VC firms are larger than two \cite{gu2019exploring}. 




\begin{figure}[!htbp]     
    \centering
    \includegraphics[scale=0.8]{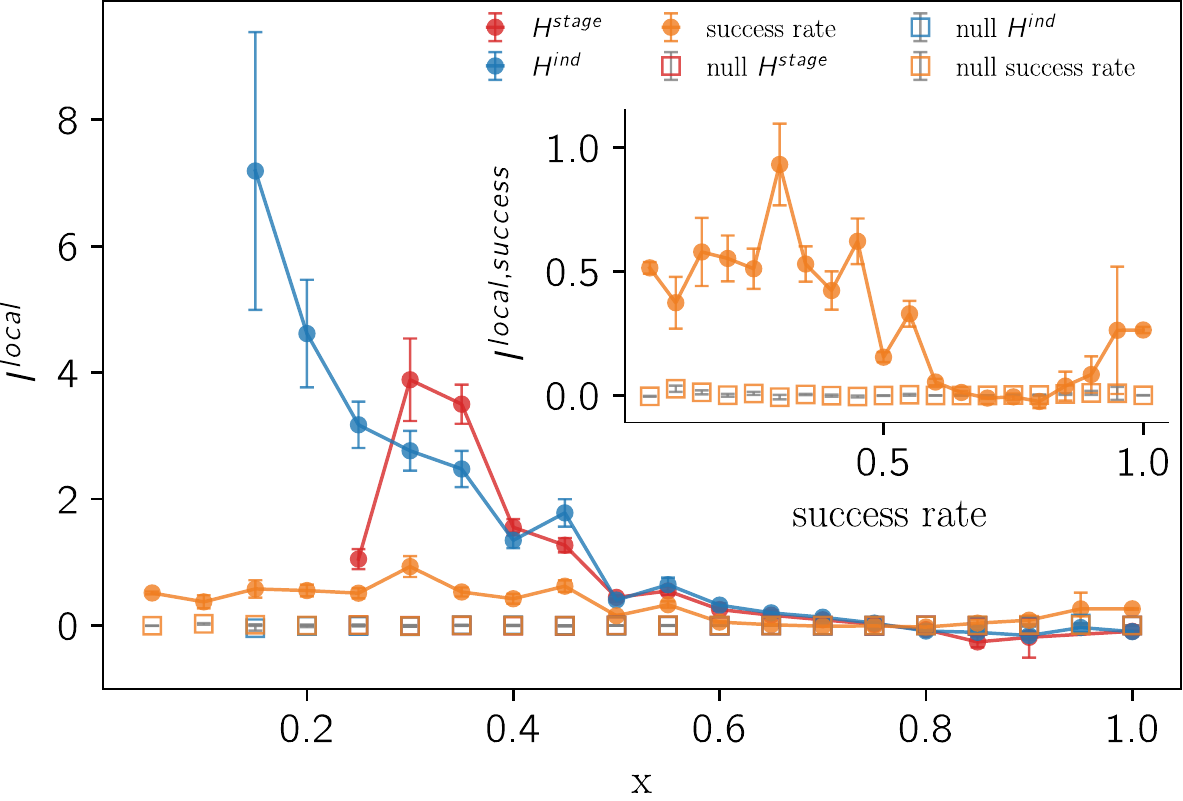}
    \caption{The relation between specialisation level ($H^{\textrm{stage}}$ (red solid circle) and $H^{\textrm{ind}}$ (blue solid circle)) or success rate (orange solid circle)) and the local Moran's I. The error bars are the standard errors of the binned values (mean values). The randomised results are shown by the open squares in the same color scheme with the grey error bars. The weight matrix $\dist_{ij}^{(1)}=\adj_{ij}$, i.e, only considering first-order neighbours. The null model is implemented as follows: for each node, fix the concerned value of it and shuffle values of remaining nodes, then recalculate the local Moran's I of this node, and this process is repeated for 200 times to obtain an average.}
    \label{fig:morani_local}
\end{figure}

However, counter-intuitively, this global gathering pattern can not represent the local similarity patterns, and the three variables ($H^{ind}$, $H^{stage}$, and success rate), distribute on the networks heterogeneously. The specialisation level ($H^{ind}$ or $H^{stage}$) or success rate of each node and their local Moran's I is shown in Fig.~\ref{fig:morani_local}. 
Given the fact that there are many small VC firms with a high specialisation level due to a limited number of investment activities (see Fig. \ref{fig.A:HHIdistribution} in Appendix), it is nontrivial to observe that specialists have small local Moran's I values, while generalists typically have larger local Moran’s I values, which indicates a strong local clustering in their neighbourhoods. 
Additionally, we can find a clear decreasing trend of averaged local Moran's I along with the increase of HHI of both industry and stage. This indicates that when a VC firm becomes more specialised, their neighbours tend to be more random, i.e., with both similar and dissimilar neighbours around. 

Compared to clustering of specialisation, the local Moran's I on success rate is not that high (see orange line in Fig. \ref{fig:morani_local}), indicating weak dependence of success level on local Moran's I. The local patterns of VC firms with higher success rate is within the error bar of the null model (except for the success rate $1$). In other words, co-investing with successful investors will not guarantee a successful investment.  
However, the averaged local Moran's I is higher than the null model for most nodes with a low success rate (see the zoom-in figure in~Fig.~\ref{fig:morani_local}). This suggests that VC firms of poor performance may have a stronger clustering, while the performance of neighbours around relative successful VC firms are more random. Our results take new measurements of Moran's I, which is not considered by the previous study, such as~\cite{abell2007performance}. Their conclusion that the better networked VC firms perform better is based on the regression of network measurements and the VC performance, which overlooked the local patterns. 

\section{Discussion}

Network science has demonstrated its advantages in gaining a better understanding and insights of various physical and social systems, including investor types in shareholding networks~\cite{yao2019network}, the life cycle of companies~\cite{west2017scale}, the efficiency of transportation systems~\cite{dong2016population,ccolak2016understanding}, spatio-temporal interaction patterns between individuals \cite{liu2022revealing}, and attraction of cities on venture capital \cite{li2021assessing}. In this work, we leverage the advantages of the network modelling approach to investigate the effect of joint-investment relationships on the specialisation and performance of VC firms. 

We first show that the number of investments made by VC firms, the number of investments received by start-up companies, and the number of IPOs all follow power-law like distributions. This fat-tail phenomenon is commonly identified in other types of real-world networks, implying collectively heterogeneous behaviours of a large number of VC firms. Then we construct a joint-investment syndication network, the nodes representing VC firms and the edges representing investments made by the connected nodes in the same start-up company at the same time. The characteristics of this network are demonstrated by the distributions of degree, H-index and community ranking. 


Moreover, the statistics of the co-investors when entering a new stage or industry shows a tendency to co-invest with new friends. However, we cannot conclude whether there is a causal relationship between making new friends and entering a new stage or new industry. We also discover a negative correlation between the specialisation level on both investment stage and industry, which are measured by HHI, and the degree of the node in the syndication network. This informs us that a well-connected VC firm is likely to have diversified investment portfolios. 
Further analysis based on our adapted Moran's I on network demonstrates a decrease in the clustering phenomenon of VC firms with similar specialisation level and success rate with the topological distance. The clustering is generally restricted within one or two step of neighbours. 
Additionally, the local Moran's I analysis reveals an unequal distribution of the clustering. Instead of observing a successful club of VC firms connected by joint-investment relationships, we find it is likely that VC firms with low success rate cluster together. 

In short, our work comprises the study of VC firms from a network perspective that appreciates the importance of interactions between entities. The analysing framework of this study can be applied to other similar scenarios to understand the network effects on the node behaviours. 

Future study can be expanded upon. In the recent decade, as the Chinese VC market becomes more mature, the managers (general partners) play a more important role in the investment. Thus, when team-level data is available, especially the social network between managers, instead of constructing the joint-investment network between VC firms, we can build the networks of managers and study the relationships between social networks and investment networks. When studying the performance of VC firms, we can further validate our argument by other types of financial performances, like the internal rate of return of the VC fund or the return of investment of each investment, when such data become available. Finally, regarding the causal inference between investment performance other behaviours, temporal analysis should be considered. 
 


\section*{References}
\bibliographystyle{unsrt}
\bibliography{VC}

\section*{Acknowledgments}
This work receives financial supports from the National Natural Science Foundation of China (Grant No. 61903020), Fundamental Research Funds for the Central Universities (Grant No. buctrc201825).

\section*{Data availability statement}
The raw data on venture capital investments is purchased from SiMuTong dataset of Zero2IPO Group (\url{www.pedata.cn}), for which we cannot disclose. 
The codes will be available upon reasonable request to the authors.

\section*{ORCID iDs}
Qing Yao \url{https://orcid.org/0000-0002-5222-8977}

\noindent
Ruiqi Li~\:\;\url{https://orcid.org/0000-0002-3290-9476}

\noindent
Kim Christensen \url{https://orcid.org/0000-0002-7648-7332}

\clearpage
\section*{Appendix}
\renewcommand\thefigure{A.\arabic{figure}}    
\setcounter{figure}{0}   
\renewcommand\thetable{A.\arabic{table}}    
\setcounter{table}{0}   

\begin{figure}[!htbp]    \centering
\includegraphics[scale=1]{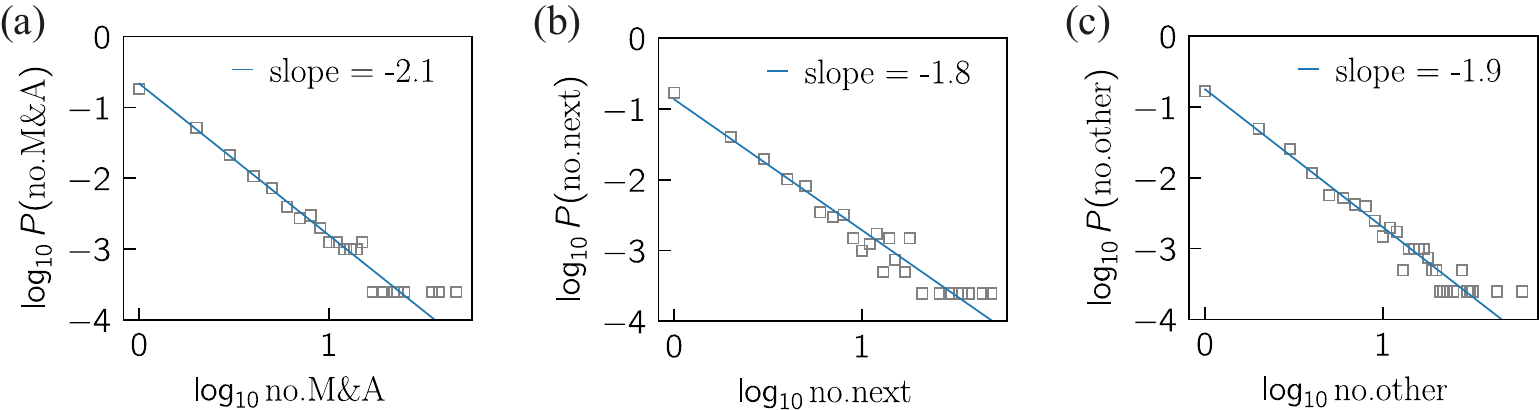}
\caption{The distribution of (a) the number of M\&A (Mergers \& Acquisitions), (b) the number of start-up companies those got next round of investment, and (c) the number other types of unsuccessful exits of each VC firm. All distributions is well approximated by a power law.} 
\label{fig:figA1}
\end{figure}

\begin{figure}[!htbp]    \centering
\includegraphics[scale=1]{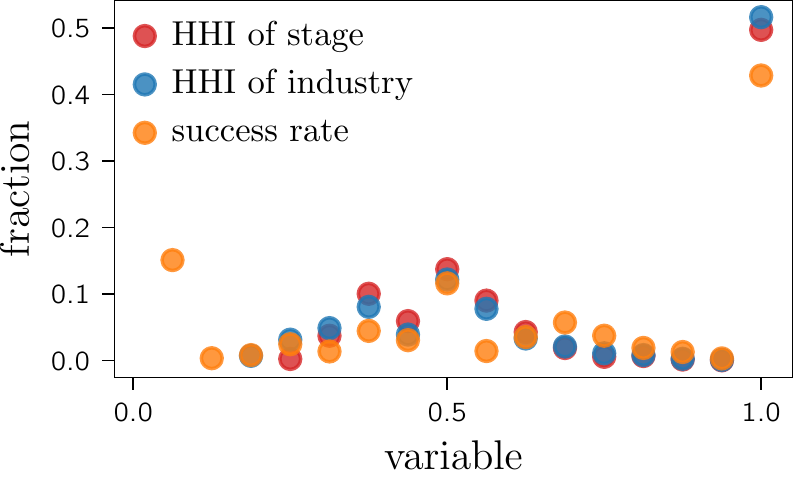}
\caption{The distribution of HHI of industry ($H^{ind}$), investment stage ($H^{stage}$), and success rate of VC firms. Due to the fact that many VC firms only make few investments (e.g., one or two investments), they are of a higher probability of having an HHI as one or success rate as one or zero.} 
\label{fig.A:HHIdistribution}
\end{figure}



\begin{table}[!htbp]
\resizebox{\textwidth}{!}{\begin{tabular}{|c|c|c|c|c|c|c|c|c|c|}
\hline
Investments & 1st & 2nd     & 3rd     & 4th         & 5th     & 6th     & 7th         & 8th     & ··· \\ \hline
Stage              & E   & I       & E       & S           & I       & M       & I           & M       & ··· \\ \hline
Industry           & A   & A       & I       & C           & C       & B       & B           & T       & ··· \\ \hline
Co-investor        &     & vc$_3$     & vc$_1$,vc$_2$ & vc$_4$,vc$_5$,vc$_6$ &         & vc$_2$,vc$_4$ & vc$_1$,vc$_2$,vc$_3$ & vc$_3$,vc$_4$ & ··· \\ \hline
Null model         & vc$_1$ & vc$_4$,vc$_2$ & vc$_2$,vc$_1$ & vc$_3$         & vc$_4$,vc$_3$ & vc$_6$,vc$_2$ & vc$_4$         & vc$_5$,vc$_3$ & ··· \\ \hline
\end{tabular}}
\caption{An example of the investment sequence of a VC firm. The investments are ordered by time. The stage corresponds to seed (S), initial (I), expansion (E), and mature (M), respectively. Letters in the Industry row corresponds to one of the $17$ industry classifications by ICNEA. }\label{tab:A.randomization}
\end{table}

\end{document}